\documentclass[aps,twocolumn]{revtex4}

\newcommand{\be}{\begin{equation}}
\newcommand{\ee}{\end{equation}}

\begin{document}

\title{Fractional Systems and Fractional Bogoliubov Hierarchy Equations
\footnote{{\it Physical Review E 71 (2005) 011102}}}

\author{Vasily E. Tarasov$^{*}$}

\affiliation{\it Skobeltsyn Institute of Nuclear Physics,
Moscow State University, Moscow 119992, Russia }

\email{tarasov@theory.sinp.msu.ru}

\begin{abstract}
We consider the fractional generalizations of the phase volume, 
volume element and Poisson brackets.
These generalizations lead us to the fractional analog of the phase space.
We consider systems on this fractional phase space
and fractional analogs of the Hamilton equations.
The fractional generalization of the average value is suggested.
The fractional analogs of the Bogoliubov hierarchy
equations are derived from the fractional Liouville equation.
We define the fractional reduced distribution functions.
The fractional analog of the Vlasov equation and
the Debye radius are considered.
\end{abstract}

\pacs{05.40.-a, 02.50.Ey, 05.20.-y}

\keywords{Fractional integrals, phase space, Poisson brackets,
chaotic dynamics, Bogoliubov equations }

\maketitle

\section{Introduction}

Derivatives and integrals of fractional order have found many
applications in recent studies of scaling phenomena \cite{1,2,3,4,MK,Zas2}.
In Ref. \cite{Zas}, coordinate fractional derivatives
in the Fokker-Planck equation were used.
It is known that Fokker-Planck equation can be derived
from the Liouville equation \cite{Is}.
The Liouville equation is derived from the
normalization condition and the Hamilton equations \cite{KSF}.
In the Hamilton equations
$dq_{k}/dt=p_{k}/m$, $dp_{k}/dt=F_k(q,p)$ 
we have only the time derivatives.
The usual normalization condition leads to the usual
(non-fractional) Liouville equation.
If we would like to derive the fractional Liouville equation
then we must use a fractional normalization condition.
In Ref. \cite{chaos}, the fractional Liouville equation was derived
from fractional normalization condition.

The natural question arises: What could be the physical
meaning of the fractional normalization condition?
This physical meaning can be the following:
the system is to be found somewhere
in the fractional phase space.
The fractional normalization condition can be considered as a
normalization condition for the distribution function in a
fractional phase space.
In order to use this interpretation we must define
a fractional phase space.
The first interpretation of the fractional phase space
is connected with fractional dimension space.
The fractional dimension interpretation follows from the
formulas for dimensional regularizations.
This interpretation was suggested in Ref. \cite{chaos}.
In this paper we consider the second interpretation of the
fractional phase space.
This interpretation follows from the fractional measure
of phase space \cite{chaos} that is used in the fractional integrals.
The fractional phase space is considered as a phase space 
that is described by the fractional powers of coordinates and momenta
($q^{\alpha}_k,p^{\alpha}_k$).
In this case, the fractional normalization condition 
for the distribution function and the fractional average values
are considered as a condition  and values 
for the fractional space.
In general, these systems are non-Hamiltonian dissipative 
systems for the usual phase space $(q,p)$. 

It is known that Bogoliubov equations can be derived from the
Liouville equation and the definition of the average value
\cite{Bog,Bog3,Gur,Petrina}.
In Ref. \cite{chaos}, the fractional Liouville equation
is derived from the fractional normalization condition.
In this paper we define the fractional analog of the average
value and reduced distributions. 
Then we derive the fractional Bogoliubov equations
from the fractional Liouville equation and the definition 
of the fractional average value.
We derive the fractional analog of the Vlasov equation 
and the Debye radius.

In Ref. \cite{chaos}, the
Riemann-Liouville definition of the fractional
integration and differentiation is used.
Therefore in this paper we use this definition
of fractional integration and differentiation.

In Sec. II, we define the fractional phase space volume. 
The fractional phase volume element for the fractional phase space
is considered.
We define the fractional analog of the Poisson bracket. 
In Sec. III, we consider the fractional systems. We discuss the
free motion of the fractional systems, the fractional harmonic oscillator
and fractional analog of the Hamiltonian systems. 
In Sec. IV,  the fractional average values and some notations
are considered. We define the reduced one-particle and two-particle
distribution functions. 
In Sec. V, the fractional Liouville equation for
$n$-particle system is written. 
We derive the first fractional Bogoliubov equation
from the fractional average value and the fractional Liouville equation.
The second fractional Bogoliubov equation is considered.
In Sec. VI, we derive the fractional analog of the Vlasov 
equation and the Debye radius for fractional systems.
Finally, a short conclusion is given in Sec. VII.

\section{Fractional Phase Volume and Poisson Brackets}

\subsection{Fractional Phase Volume for Configuration Space}

Let us consider the phase volume for the region 
such that $x\in[a;b]$.
The usual phase volume is defined by
\be \label{V1} V_1 =\int^b_a dx =\int^y_a dx +\int^b_y dx  . \ee
The fractional integrations are defined \cite{SKM} by
\[ I^{\alpha}_{a+}1=
\frac{1}{\Gamma (\alpha)} \int^{y}_{a} \frac{dx}{(y-x)^{1-\alpha}}, \]
\[ I^{\alpha}_{b-}1=
\frac{1}{\Gamma (\alpha)} \int^{b}_{y} \frac{dx}{(x-y)^{1-\alpha}}. \]
Using these notations, we get Eq. (\ref{V1}) in the form
\be \label{V2} V_1 =I^{1}_{a+}1+I^{1}_{b-}1 . \ee
The fractional generalization of this phase volume can be defined by
\be \label{Valpha} V_{\alpha}=I^{\alpha}_{a+}1+I^{\alpha}_{b-}1 . \ee

The fractional phase volume integral (\ref{Valpha}) 
can be written in the form
\be \label{V+} V_{\alpha}= \frac{1+\lambda^{\alpha}}{\Gamma(\alpha)}
\int^{b-y}_{0} \xi^{\alpha-1} d\xi  \ee
where $\lambda$ is defined by
\[ \lambda =\frac{y-a}{b-y} . \]

In order to have the symmetric limits of the 
phase volume integral we can use the following equation: 
\be \label{PV4} V_{\alpha}=\frac{(1+\lambda^{\alpha})}{2}
\int^{+(b-y)}_{-(b-y)} d \mu_{\alpha}(x) , \ee
where
\be d \mu_{\alpha}(x)=\frac{|x|^{\alpha-1}dx}{\Gamma(\alpha)}=
\frac{dx^{\alpha}}{\alpha \Gamma(\alpha)} . \ee
Here we use the following notation for fractional power of
coordinates:
\be \label{xa}
x^{\alpha} =\beta(x) (x)^{\alpha}= sgn(x) |x|^{\alpha} , \ee
where $\beta(x)=[sgn(x)]^{\alpha-1}$. 
The function $sgn(x)$ is equal to $+1$ for $x\ge0$,
and this function is equal $-1$ for $x<0$.

\subsection{Fractional Phase Volume for Phase Space}

Using Eq. (\ref{PV4}), we have the phase volume for
the two-dimensional phase space in the form
\be \label{PV5} V_{\alpha}= \frac{(1+\lambda^{\alpha}_q)}{2}
\frac{(1+\lambda^{\alpha}_p)}{2} \int^{+(q_b-y)}_{-(q_b-y')}
\int^{+(p_b-y)}_{-(p_b-y')}
\frac{dq^{\alpha} \wedge d p^{\alpha}}{(\alpha \Gamma(\alpha))^{2}} \ee
where
\be \label{muV0} dq^{\alpha} \wedge d p^{\alpha} =\alpha^2
|q p|^{\alpha-1} dq \wedge d p . \ee

The fractional measure for the region $B$ of 2n-dimensional phase 
space can be defined by the equation
\be  \label{sim1} \mu_{\alpha}(B)=V_{\alpha}=
\int_B g(\alpha) d\mu_{\alpha}(q,p), \ee
where $d\mu_{\alpha}(q,p)$ is a phase volume element, 
\be \label{muV} d\mu_{\alpha}(q,p)=
\prod^n_{k=1} \frac{dq^{\alpha}_k \wedge d p^{\alpha}_k}{
(\alpha \Gamma(\alpha))^{2}} , \ee
and $g(\alpha)$ is a numerical multiplier, 
\[ g(\alpha)=\frac{1}{4^{n}} \prod^{n}_{k=1} g_k(\alpha) . \]
If the domain $B$ of the phase space is defined by
$q_k\in R^1$ and $p_k\in R^1$, then $g_k(\alpha)=1$.
If this domain is defined by $q_k\in [q_{ak};q_{bk}]$
and $p_k\in [p_{ak};p_{bk}]$, then
\be g_k(\alpha)=
\Bigl(1+\Bigl(\frac{q_{bk}-y_k}{y_k-q_{ak}}\Bigr)^{\alpha}\Bigr)
\Bigl(1+\Bigl(\frac{p_{bk}-y'_k}{y'_k-p_{ak}}\Bigr)^{\alpha}\Bigr) \ee
It is easy to see that the fractional measure depends on the fractional
powers $(q^{\alpha},p^{\alpha})$.

\subsection{Fractional Exterior Derivatives}

In Eq. (\ref{PV5}), we use the usual exterior derivative
\be \label{d0} d=\sum^{2n}_{k=1} dx_k \frac{\partial}{\partial x_k} ,\ee
Obviously, this derivative can be represented in the form
\be \label{14} d=\sum^{2n}_{k=1} dx^{\alpha}_k
\frac{\partial}{\partial x^{\alpha}_k} , \ee
where $x^{\alpha}$ is defined by Eq. (\ref{xa}).

Note that the volume element of fractional phase space can be realized 
by fractional exterior derivatives \cite{CN}, 
\[ d^{\alpha}=\sum^{n}_{k=1}dq^{\alpha}_k \frac{\partial^{\alpha}}{(\partial 
(q_k-y_k))^{\alpha}}+\sum^{n}_{k=1}
dp^{\alpha}_k \frac{\partial^{\alpha}}{(\partial (p_k-{y'}_k))^{\alpha}}, \]
in the following form:
\[ dq^{\alpha} \wedge dp^{\alpha}= \Bigl(\frac{4}{\Gamma^2(2-\alpha)}+
\frac{1}{\Gamma^2(1-\alpha)}\Bigr)^{-1} 
(qp)^{\alpha-1} d^{\alpha}q \wedge d^{\alpha}p . \]

\subsection{Fractional Poisson Brackets}

We can define the fractional generalization of the 
Poisson brackets in the form
\be \label{PB2} \{A,B\}_{(\alpha)}=\sum^n_{k=1}\Bigl(
\frac{\partial^{\alpha} A}{(\partial q_k)^{\alpha}}
\frac{\partial^{\alpha} B}{(\partial p_k)^{\alpha}}-
\frac{\partial^{\alpha} A}{(\partial p_k)^{\alpha}}
\frac{\partial^{\alpha} B}{(\partial q_k)^{\alpha}} \Bigr) , \ee
where we use the fractional derivatives \cite{SKM}. 
It is known  \cite{SKM} that 
the derivative of a constant need not be zero, 
\be \label{1ne0} \frac{\partial^{\alpha} 1}{(\partial x)^{\alpha}}=
\frac{1}{\Gamma(1-\alpha)} x^{-\alpha} .\ee
This equation leads us to the correlation between coordinates $q$
and momenta $p$ in the form
\be \label{qp1} \frac{\partial^{\alpha} p}{(\partial q)^{\alpha}}=
\frac{pq^{-\alpha}}{\Gamma(1-\alpha)} , \quad
\frac{\partial^{\alpha} q}{(\partial p)^{\alpha}}=
\frac{qp^{-\alpha}}{\Gamma(1-\alpha)}  . \ee
It is easy to see that $q$ and $p$ are not independent 
variables in the usual sense.
Therefore the fractional analog (\ref{PB2})
of the Poisson brackets is not convenient.

In the general case ($n\ge 2$), the Poisson brackets
$\{q_k,q_l\}_{(\alpha)}$ and $\{p_k,p_l\}_{(\alpha)}$
(with $k\not=l$) are not equal to zero.
If we consider $n=2$, then we have
\be \{q_1,q_2\}_{(\alpha)}= -\gamma_2(\alpha) q_1 q_2
[ (q_1p_1)^{-\alpha}-(q_2 p_2)^{-\alpha}] ,\ee
where the coefficient $\gamma_2(\alpha)$ is defined
by the equation
\be \gamma_2(\alpha)=\frac{1}{\Gamma^2(1-\alpha)}-
\frac{2}{\Gamma(1-\alpha) \Gamma (2-\alpha)}  . \ee
Moreover, the Poisson brackets with the unit are not equal to zero.
Using Eq. (\ref{1ne0}), we get
\be \{1,q\}_{(\alpha)}= \gamma_2(\alpha) q^{1-\alpha} p^{\alpha}, \ee
\be \{1,p\}_{(\alpha)}=-\gamma_2(\alpha) q^{-\alpha} p^{1-\alpha} . \ee
Therefore the fractional Poisson brackets (\ref{PB2}) are not convenient.

We can use the fractional power Poisson brackets:
\be \label{PB} \{A,B\}_{\alpha}=
\sum^n_{k=1}\Bigl(\frac{\partial A}{\partial q^{\alpha}_{k}}
\frac{\partial B}{\partial p^{\alpha}_{k}}-
\frac{\partial A}{\partial p^{\alpha}_{k}}
\frac{\partial B}{\partial q^{\alpha}_{k}} \Bigr) , \ee
where we use notations (\ref{xa}).
Obviously, the relations
\[ \{q^{\alpha}_{k},q^{\alpha}_{l} \}_{\alpha}=0, \quad
\{p^{\alpha}_{k},p^{\alpha}_{l} \}_{\alpha}=0 , \quad
\{q^{\alpha}_{k},p^{\alpha}_{l} \}_{\alpha}=\delta_{kl} , \]
are realized for these Poisson brackets.
Therefore the fractional Poisson brackets (\ref{PB}) 
are more convenient.

\subsection{Fractal Dimension of Space}

The interpretation of the fractional phase space
is connected with the fractional measure of phase space.
The parameter $\alpha$  defines the space with
the fractional phase measure (\ref{sim1}) and (\ref{muV}). 
It is easy to prove that the velocity of the fractional 
phase volume change is defined by the equation
\[ \frac{dV_{\alpha}}{dt}=
\int_B \Omega_{\alpha}(q,p,t) g(\alpha) d\mu_{\alpha}(q,p) ,\]
where  
\be \label{Oa} \Omega_{\alpha}=\{\frac{dq^{\alpha}_t}{dt},p^{\alpha}_t\}_{\alpha}+
\{q^{\alpha}_t,\frac{dp^{\alpha}_t}{dt}\}_{\alpha} . \ee
Equation. (\ref{Oa}) is proved in Ref. \cite{chaos}.
The form of the omega function allows us to consider 
a different class of the systems that are described by 
the fractional powers of coordinates and momenta.

The interpretation of the fractional phase space  
can be derived from the fractional measure 
that is used in the fractional integrals. 
The interpretation of the fractional phase space
can be connected with fractional dimension.
We have two arguments for this point of view.

1. Let us use the well-known formulas for dimensional 
regularizations \cite{Col}:
\be \label{dr} \int \rho(x) d^{n} x =
\frac{2 \pi^{n/2}}{\Gamma(n/2)}
\int^{\infty}_{0} \rho(x)  x^{n-1} dx  . \ee
Using Eq. (\ref{dr}), we get that the fractional integral can be 
considered \cite{chaos} as an integral for the fractional dimension space
\be \label{fnc-2} \frac{\Gamma(\alpha/2) }{2 \pi^{\alpha/2} \Gamma(\alpha)}
\int \tilde \rho (x,t) d^{\alpha} x  \ee
up to the numerical factor
$\Gamma(\alpha/2) /( 2 \pi^{\alpha/2} \Gamma(\alpha))$.

2. Let us consider the well-known definition of the fractal mass 
dimension.
The equations that define the fractal dimensions have 
the passage to the limit. This passage makes difficult 
the practical application to the real fractal media.
The other dimensions, which can be easily calculated from
the experimental data, are used in the empirical investigations.
For example, the mass fractal dimension \cite{Mand,Schr} 
can be easy measured.

The properties of the fractal media like mass obeys a 
power law relation, 
\be \label{MR} M(r) =kr^{D_m} , \ee
where $M$ is the mass of fractal medium, $r$ is a box size 
(or a sphere radius), and $D_m$ is a mass fractal dimension. 
The amount of mass of a medium inside a box of size $r$
has a power law relation (\ref{MR}).

The fractal dimension can be calculated by box counting method
which means drawing a box of size r 
and counting the mass inside. 
To calculate the mass fractal dimension,  take the 
logarithm  of both sides of Eq. (\ref{MR}):
\[ ln (M)=D_m \ ln(r)+ln k . \]
The log-log plot of $M$ and $r$ gives us the slope $D_m$, 
the fractal dimension. 
When we graph ln(M) as a function of ln(r), we get a value 
of about $D_m$ which is the fractal dimension of fractal media.

The power law relation (\ref{MR}) can be naturally 
derived by using the fractional integral.
In order to describe the fractal media, we suggest to use 
the space with fractional measure. 

Let us consider the line distribution of the mass.
If we consider the mass of the homogeneous distribution 
($\rho=const$) in the ball region $W$ with radius $r$, then we have
\be \label{m1} M_1(r)=\int^{+r}_{-r} \rho(x) dx= 2 \rho
\int^{r}_{0} dx= 2 r^1 . \ee
In this case, $D_m=1$.
Let us consider line mass distribution in the fractional space.
In that case, a ball of radius $r$ covers a mass 
\[ M_{\alpha}(r)=\frac{2 \rho r^{\alpha}}{\alpha \Gamma(\alpha)}. \]
This equation can be proved by the fractional generalization 
of Eq. (\ref{m1}) in the form
\be M_{\alpha}(r)=\frac{2}{\Gamma(\alpha)}
\int^{r}_{0} \rho(\xi) \xi^{\alpha-1} d\xi =\rho V_{\alpha} (r)=
\frac{2 \rho r^{\alpha}}{\alpha \Gamma(\alpha)} . \ee
The initial points in the fractional integral (\ref{V+}) 
are set to zero, and $a=-r$, $b=r$.
The fractal dimension of particle system and fractal media 
is defined as the exponent of $r$ in the growth law for 
mass $M(r)$ or number of particles,  
\be \label{Nr} n(r)=M(r)/m=(k/m)r^{D_m} . \ee
Here $m$ is a particle mass.
Thus we see that the fractal dimension $D_m$ of particle system 
(in the fractional space) is $\alpha$, i.e., $D_m=\alpha$. 
Therefore the space with fractional measure (\ref{V+}) 
can be considered as a space with fractal dimension $D_m=\alpha$. 

As the result the space with fractional measure (\ref{V+}) can be 
used to describe the particle systems and medium with 
non-integer mass dimension.

\section{Fractional systems}

\subsection{Equations of Motion}

In Sec. II we prove that the fractional measure (\ref{muV})
depends on the fractional powers $(q^{\alpha},p^{\alpha})$, 
and Poisson brackets (\ref{PB}) with fractional powers are 
more convenient. Therefore we can consider a different class of mechanical 
systems that are described by the fractional powers of coordinates 
and momenta.
We can consider the fractional power of the coordinates 
as a convenient way to describe systems in the fractional dimension space.

Let us consider a classical system with the mass $M$. 
Suppose this system is described by the dimensional
coordinates ${\bar q}_k$ and momenta ${\bar p}_k$ 
that satisfy the following equations of motion:  
\be \label{bar1} \frac{d {\bar q}_k}{d {\bar t}}=\frac{{\bar p}_k}{M} ,
\quad
\frac{d {\bar q}_k}{d {\bar t}}=f_k({\bar q},{\bar p},{\bar t}) . \ee
Let $q_0$ be the characteristic scale in the configuration space;
$p_0$ be the characteristic momentum, $F_0$ be the characteristic value
of the force, and $t_0$ be the typical time.
Let us introduce the {\it dimensionless variables}
\[ q_k=\frac{{\bar q}_k}{q_0}, \quad p_k=\frac{{\bar p}_k}{p_0}, \quad
t=\frac{{\bar t}}{t_0}, \quad F_k=\frac{f_k}{F_0} . \]
Here and later we use $q_k$ and $p_k$  as dimensionless variables. 
Using Eq. (\ref{bar1}) for dimensional physical variables $({\bar q},{\bar p})$,
we get the equations for dimensionless variables,  
\be \label{bar2} \frac{d q_k}{d t}=\frac{p_k}{m} , \quad
\frac{d q_k}{d {\bar t}}=A F_k(q,p,t) , \ee
where $m=Mq_0/t_0p_0$ is the dimensionless mass, and
\be \label{A1} A=\frac{t_0 F_0}{p_0}. \ee
Using the dimensionless variables $(q,p,t)$, we can consider the fractional 
generalization of Eq. (\ref{bar2}) in the form
\be \label{bar3} \frac{d q^{\alpha}_k}{d t}=\frac{p^{\alpha}_k}{m} , 
\quad
\frac{d q^{\alpha}_k}{d {\bar t}}=A F_k(q^{\alpha},p^{\alpha},t) , \ee
where we use the following notations: 
\be \label{qa} q^{\alpha}_k =\beta(q) (q_k)^{\alpha}= sgn(q_k) |q_k|^{\alpha}, \ee
\be \label{pa} p^{\alpha}_k =\beta(p) (p_k)^{\alpha}= sgn(p_k) |p_k|^{\alpha}. \ee
Here $k=1,...,n$, and $\beta(x)$ is defined by Eq. (\ref{xa}).

A system is called a fractional system if the phase space of the system
can be described by the fractional powers of coordinates (\ref{qa}) 
and momenta (\ref{pa}).

The fractional phase space can be considered as a 
phase space for the fractional systems.
This interpretation follows from the fractional measure that
is used in the fractional integrals. 

We can consider the fractional systems in the usual phase space $(q,p)$
and in the fractional phase space $(q^{\alpha},p^{\alpha})$.
In the second case, the equations of motion for the fractional 
systems have more simple form. 
Therefore we use the fractional phase space.
The fractional space is considered as a space with the fractional measure 
that is used in the fractional integrals. 

The fractional generalization of the conservative Hamiltonian system 
is described by the equation
\be \label{33} \frac{dq^{\alpha}_{k}}{dt}=
\frac{\partial H}{\partial p^{\alpha}_{k}},
\quad \frac{dp^{\alpha}_{k}}{dt}=
-\frac{\partial H}{\partial q^{\alpha}_{k}}, \ee
where $H$ is a fractional analog of the Hamiltonian.
Note that the function $H$ is the invariant of the motion. 
Using the fractional Poisson brackets (\ref{PB}), we have 
\be \frac{dq^{\alpha}_{k}}{dt}=\{q^{\alpha}_{k},H\}_{\alpha},
\quad \frac{dp^{\alpha}_{k}}{dt}=\{p^{\alpha}_{k},H\}_{\alpha}. \ee
Here we use Poisson brackets (\ref{PB}). 
These equations describe the system in the fractional phase space 
$(q^{\alpha},p^{\alpha})$.  
For the usual phase space $(q,p)$, the fractional Hamiltonian systems
are described by the equations
\be \label{35} \frac{dq_{k}}{dt}=\frac{(q_kp_k)^{1-\alpha}}{\alpha^2}
\frac{\partial H}{\partial p_{k}},
\quad \frac{dp_{k}}{dt}=- \frac{(q_kp_k)^{1-\alpha}}{\alpha^2}
\frac{\partial H}{\partial q_{k}}. \ee
The fractional Hamiltonian systems are non-Hamiltonian
systems in the usual phase space $(q,p)$. 
A classical system (in the usual phase space) 
is called Hamiltonian if the right-hand sides of the equations
\be \label{qg-pf} \frac{dq_{k}}{dt}=g_k(q,p),
\quad \frac{dp_{k}}{dt}=f_k(q,p) \ee
satisfy the following Helmholtz conditions \cite{Tar-tmf3}:
\be
\frac{\partial g_k}{\partial p_l}-\frac{\partial g_l}{\partial p_k}=0,
\quad
\frac{\partial g_k}{\partial q_l}-\frac{\partial f_l}{\partial p_k}=0,
\quad
\frac{\partial f_k}{\partial q_l}-\frac{\partial f_l}{\partial q_k}=0.
\ee
It is easy to prove that the Helmholtz conditions are not satisfied. 
Therefore the fractional Hamiltonian system (\ref{35})
is a non-Hamiltonian system in the usual phase space $(q,p)$.
The fractional phase space allows us to write the equations 
of motion for the fractional systems in the simple form (\ref{33}).

If we have $dq^{\alpha}_k/dt=p^{\beta}_k/m$, then the fractional 
analog of the Hamiltonian can be considered in the form 
\be \label{Hb} H=
\sum^n_{k,l=1} \frac{\alpha p^{\alpha+\beta}_{k}}{m(\alpha + \beta)}
+U(q) . \ee
 
If $\beta=\alpha$, then
we have the fractional analog of the kinetic energy $T=p^{2\alpha}/2m$.

The omega function for the system (\ref{qg-pf}) 
in the usual phase space $(q,p)$
is defined by the equation
\be \label{51}
\Omega=\sum^n_{k=1}\Bigl( \frac{\partial g_k}{\partial q_k}+
\frac{\partial f_k}{\partial p_k}\Bigr) .
\ee
If the omega function is negative $\Omega<0$, then 
the system is called a dissipative system.
If $\Omega \not=0$, then the system is a generalized
dissipative system.
For the fractional Hamiltonian systems (\ref{35}), 
the omega function (\ref{51}) is not equal to zero.
Therefore the fractional Hamiltonian systems are the 
general dissipative systems in the usual phase space.

It is known that the non-Hamiltonian and dissipative systems 
(for example, $F=-\gamma p$) are not invariant 
under the Galilean transformation.
In general, the fractional systems are non-Hamiltonian systems
for the usual phase space $(q,p)$. 
Therefore the Galilean transformations for the
equations of motion are not considered in this paper. 
The fractional analogs of the Hamiltonian systems can be invariant
under the fractional generalization of the Galilean transformation.

It is not hard to prove that the fractional systems (\ref{Hb})
are connected with the non-Gaussian statistics.
Classical dissipative and non-Hamiltonian systems can have
the canonical Gibbs distribution as a solution of the stationary
Liouville equations  \cite{Tar-mplb}.
Using the methods \cite{Tar-mplb}, it is easy to prove
that some of fractional dissipative systems can have
fractional Gibbs distribution (non-Gaussian statistic)
\be \rho(q,p)=exp \ [{\cal F}-H(q,p)]/kT, \ee
as a solution of the fractional Liouville equations \cite{chaos}.
The interest in and relevance of fractional kinetic equations
is a natural consequence of the realization of the importance of
non-Gaussian statistics of many dynamical systems. There is already
a substantial literature studying such equations in one or more
space dimensions.

\subsection{Free Motion of Fractional System}

Let us consider the free motion of the fractional system 
that is defined by the following equations: 
\be \label{free1}
\frac{dq^{\alpha}}{dt}=\frac{p^{\alpha}}{m} ,\quad
\frac{dp^{\alpha}}{dt}=0 .
\ee
The solutions of these equations with the conditions
$q(0)=q_1$ and $p(0)=p_1$ have the form
\be \label{solu}
q^{\alpha}(t)=m^{-1}p^{\alpha}_1 t+q^{\alpha}_1, 
\quad p(t)=p_1.
\ee
We can conclude that the free motion of the fractional system
is described by 
\be
q(t)\sim (t-t_1)^{1/\alpha} ,
\ee
where the parameter $t_1$ is defined by $q_1$.

Let us consider the special cases of the parameter $\alpha$:
$0.5$, $1/3$, $0.6$.

{\bf 1}. If the parameter $\alpha$ is equal to $0.5$, then we have
the solution of Eq. (\ref{free1}) in the form
\be
q(t)=z(q_1,p_1,t) (at^2+bt+c),
\ee
where we use the following notations
\be
z(q_1,p_1,t)=sgn(\sqrt{|p_1|}t-m\sqrt{|q_1|}) , 
\ee
\be
a=|p_1|/m, \quad b=2\sqrt{|q_1p_1|}/m, \quad c=|q_1| .
\ee

{\bf 2}. If the parameter $\alpha$ is equal to $1/3$, then we have
the solution of Eq. (\ref{free1}) in the form
\be
q(t)=\frac{p_1}{m^3} t^3+\frac{3p^{2/3}_1q^{1/3}_1}{m^2} t^2
+\frac{3p^{1/3}_1q^{2/3}_1}{m} t+q_1.
\ee

{\bf 3}. If the parameter $\alpha$ is equal to $0.6$, then we have
the solution of Eq. (\ref{free1}) in the form
\be
q(t)=\Bigl(\frac{p^{3/5}_1}{m} t+q^{3/5}_1\Bigl)^{5/3}.
\ee
If $q_1=0$, then $q(t) \sim t^{5/3}$.

For the usual phase space $(q,p)$, the equations 
of motion for the fractional system 
can be represented in the following form (\ref{qg-pf}). 
For the free fractional system, we have
\be
\frac{dq_k}{dt}=\frac{q^{1-\alpha}p^{\alpha}}{\alpha m} , 
\quad \frac{dp_k}{dt}=0.
\ee
The omega function for the free fractional system
is equal to the following function:
\be
\Omega=\frac{1-\alpha}{\alpha m} q^{-\alpha} p^{\alpha}.
\ee
In general, this function does not equal to zero and the
phase volume of the usual phase space changes.
Using Eq. (\ref{solu}), we get
\be
\Omega=\frac{1-\alpha}{\alpha} (t+t_1)^{-1} ,
\ee
where $t_1=mq^{\alpha}_1/p^{\alpha}_1$. 
If $q_1=0$, then the omega function is proportional to $1/t$.
Therefore the velocity of elementary phase volume
change for the free motion in the usual phase space 
$(q,p)$ is inversely proportional to the time.  
We suppose that the initial momentum is not equal 
to zero $p_1\not=0$.

For the fractional phase space $(q^{\alpha},p^{\alpha})$,
we define \cite{chaos} the omega function $\Omega_{\alpha}$,  
in Eq. (\ref{Oa}). This "fractional" omega function 
is equal to zero for the free motion of fractional system.
Therefore the fractional phase space $(q^{\alpha},p^{\alpha})$ 
is more convenient than usual phase space $(q,p)$.

\subsection{Fractional Harmonic Oscillator}

Let us consider the fractional harmonic oscillator, 
which is defined in Ref. \cite{chaos} by the equations
\be \label{ho1}
\frac{dq^{\alpha}}{dt}=\frac{p^{\alpha}}{m},
\quad
\frac{dp^{\alpha}}{dt}=-m \omega^2 q^{\alpha},
\ee
where $\omega$ is a dimensionless variable.
The solutions of these equations of motion have the form
\be \label{Sol-ho}
q^{\alpha}(t)=a \ cos(\omega t +\varphi),
\quad
p^{\alpha}(t)=-m\omega a \ sin(\omega t +\varphi),
\ee
where parameters $a, \varphi$ are defined by
\be
a=\sqrt{ q^{2\alpha}_1+\frac{p^{2\alpha}_1}{\omega^2 m^2}},
\quad
tg \varphi =-\frac{p^{\alpha}_1}{m \omega q^{\alpha}_1} .
\ee
If $\alpha=0.5$, then we have
\be
q(t)=sgn\Bigl( cos(\omega t +\varphi_1) \Bigr)
 \ a^2 \ cos^2(\omega t +\varphi). 
\ee
If $\alpha=1/3$, then we have
$q(t)= a^3 \ cos^3(\omega t +\varphi)$. 
The fractional harmonic oscillator has the following integral of motion: 
\be \label{Hho}
H=\frac{p^{2\alpha}}{2m}+\frac{m\omega^2 q^{2\alpha}}{2}=const.
\ee
This function can be considered as a fractional analog of 
Hamiltonian for Eq. (\ref{33}).

For the fractional harmonic oscillator the
function $\Omega_{\alpha}$ is equal to zero 
in the fractional phase space $(q^{\alpha},p^{\alpha})$.
Therefore this system is a conservative Hamiltonian 
non-dissipative system in the fractional space.
{\it If we use the usual phase space to describe the
fractional harmonic oscillator, then this system is 
conservative non-Hamiltonian dissipative system.}
Note that the conservative non-Hamiltonian systems
are considered in Ref. \cite{Sergi1}.

In the usual phase space $(q,p)$ the equations of motion 
for fractional oscillator have the form
\be \label{ho2}
\frac{dq}{dt}=\frac{1}{\alpha m} q^{1-\alpha} p^{\alpha},
\quad
\frac{dp}{dt}=-\frac{m\omega^2}{\alpha}  q^{\alpha} p^{1-\alpha}.
\ee
The omega function for the usual phase space $(q,p)$
is defined by Eq. (\ref{51}) in the form
\be \label{Om-ho}
\Omega=\frac{1-\alpha}{\alpha m}
q^{-\alpha} p^{-\alpha}
(p^{2\alpha}-m^2\omega^2 q^{2\alpha}) .
\ee
The elementary phase volume of the usual phase space
changes. The velocity of this change is equal to the omega function.
Substituting the solution (\ref{Sol-ho}) into Eq. (\ref{Om-ho}), 
we have
\be
\Omega=\frac{1-\alpha}{\alpha }
2  \omega  \ cot \ 2(\omega t+\varphi) ,
\ee
where we use $sin^2 \beta-cos^2 \beta=-cos 2 \beta$ and
$cot \beta=cos \beta / sin \beta$.
Therefore the fractional harmonic oscillator is a general
dissipative system in the usual phase space $(q,p)$.

For the fractional phase space $(q^{\alpha},p^{\alpha})$,
the fractional harmonic oscillator is conservative 
nondissipative Hamiltonian system. 
Therefore the fractional phase space $(q^{\alpha},p^{\alpha})$ 
is more convenient than usual phase space $(q,p)$.

The question arises: What are the fundamentals 
(different from Hamilton principle), which can lead 
to the system of dynamic equations (\ref{ho2})?
For the fractal media the harmonic oscillator 
is defined by Eq. (\ref{ho1}). 
For the usual phase space $(q,p)$, 
this equation has form (\ref{ho2}). 
Therefore the system (\ref{ho2}) can be considered as 
a system 
\be \label{Mqp}
\frac{dq}{dt}=\frac{p}{M}, \quad \frac{dp}{dt}=- M(\omega / \alpha)^2 q, \ee
where the dimensionless mass $M(q,p)=m \alpha (p/q)^{1-\alpha}$
satisfies the scaling relation 
\[ M(\lambda_1 q, \lambda_2 p)=(\lambda_2/\lambda_1)^{1-\alpha} M(q,p). \]
This property can be formulated in the following form.
If we consider the scale transformation of the characteristic values 
in the form 
\[ q_0 \ \rightarrow \ q_0/ \lambda_1 ,\quad 
p_0 \ \rightarrow \ p_0 / \lambda_2 , \]
then we have transformation of the mass 
\[ M \ \rightarrow \ (\lambda_2/\lambda_1)^{1-\alpha} M .\]
In the general case, this scaling law can be described by the
renormalization group approach \cite{BogS}.

Note that the system (\ref{Mqp}) is non-Hamiltonian system.
We consider the fractional phase space form (\ref{ho1}) 
of Eq. (\ref{ho2}) as a more fundamental. 
The fractional harmonic oscillator
is an oscillator in the fractional phase space that can be 
considered as a fractal medium.
Therefore the fractional oscillator can be interpreted 
as an elementary excitation of some fractal medium 
with noninteger mass dimension.

\subsection{Curved Phase Space}

The fractional system with the fractional analog (\ref{Hb}) 
of the Hamiltonian can be considered as a nonlinear system with
\be \label{Hqp} H(q^{\alpha},p^{\alpha})=
\sum^n_{k=1}\frac{1}{2} g_{kl}(q,p) p_k p_l +U(q) .\ee
Note that this fractional  Hamiltonian (\ref{Hqp}) defines a
nonlinear one-dimensional sigma-model \cite{Tar-mpla,Tar-pl}
in the curved phase space with metric
$g_{kl}(q,p)= m^{-1} p^{\alpha+\beta-2}_k \delta_{kl}$. 
This means that we use the curved phase space. 
Note that this fractional Hamiltonian is used in equations of motion 
(\ref{35}) that define the non-Hamiltonian flow in the usual phase space.

The curved phase space is used in the Tuckerman approach to 
the non-Hamiltonian statistical mechanics
\cite{Tuk1,Tuk2,Tuk3,Ram1,Ram2,Sergi1,Sergi2,Sergi3,Ezra1,Ezra2}. 
In their approach the suggested invariant phase space measure of 
non-Hamiltonian systems is connected with the metric of the curved phase space. 
This metric defines the generalization of the Poisson brackets. 
The generalized (non-Hamiltonian) bracket is suggested
in Refs. \cite{Sergi2, Sergi3}.
For these brackets, the Jacobi relations will not be satisfied. 
This requires the application of non-Lie algebras
(in which the Jacobi identity does not hold) and analytic quasi-groups
(which are nonassociative generalizations of groups). 
In the paper \cite{TMP4}, 
we show that the analogues of Lie algebras and groups 
for non-Hamiltonian systems are Valya algebras 
(anticommutative algebras whose commutants are Lie subalgebras)
and analytic commutant-associative loops (whose commutants are associative
subloops (groups)). 
Unfortunately, non-Lie algebras and its representations 
have not been thoroughly studied. 
Nevertheless the Riemannian treatment of the phase space is very interesting.
This approach allows us to consider the connection between the fractal dimensional 
phase space and non-Lie algebras of the vector fields in the space.

\section{Fractional Average Values and Reduced Distributions}

\subsection{Fractional Average Values for Configuration Space}

Let us derive the fractional generalization of the equation
that defines the average value of the classical observable $A(q,p)$.

The usual average value for the configuration space
\be <A>_1= \int^{\infty}_{-\infty} A(x)\rho(x) dx  \ee
can be written in the form
\be \label{If}
<A>_1=\int^y_{-\infty} A(x)\rho(x) dx +\int^{\infty}_y A(x)\rho(x) dx . \ee
Using the notations
\[ (I^{\alpha}_{+}f)(y)=
\frac{1}{\Gamma (\alpha)} \int^{y}_{-\infty}
\frac{f(x)dx}{(y-x)^{1-\alpha}} , \]
\[ \label{I-} (I^{\alpha}_{-}f)(y)=
\frac{1}{\Gamma (\alpha)} \int^{\infty}_{y}
\frac{f(x)dx}{(x-y)^{1-\alpha}} , \]
average value (\ref{If}) can be rewritten in the form
\[ <A>_1=(I^{1}_{+}A\rho)(y)+(I^{1}_{-}A\rho)(y) . \]
The fractional generalization of
this equation is defined by
\be \label{Aa} <A>_{\alpha}=
(I^{\alpha}_{+}A\rho)(y)+(I^{\alpha}_{-}A\rho)(y) . \ee
We can rewrite Eq. (\ref{Aa}) in the form
\be \label{FI5} <A>_{\alpha}= \frac{1}{2}
\int^{\infty}_{-\infty} ( (A\rho)(y-x)+ (A\rho)(y+x)) d\mu_{\alpha}(x) , \ee
where we use
\be \label{dm}
d\mu_{\alpha}(x)=\frac{|x|^{\alpha-1} dx}{\Gamma(\alpha)}=
\frac{d x^{\alpha}}{\alpha \Gamma(\alpha)} ,\ee
and $x^{\alpha}$ is defined by Eq. (\ref{xa}).

\subsection{Fractional Average Values for Phase Space}

Let us introduce some notations to consider the fractional
average value for phase space.

1. Let us define the tilde operators $T_{x_k}$ and $T[k]$.
The operator $T_{x_k}$ is defined by
\[ T_{x_k} f(...,x_k,...)
=\frac{1}{2}\Bigl( f(...,x^{\prime}_k-x_k,...)
+f(...,x^{\prime}_k+x_k,...) \Bigr) . \]
This operator allows us to rewrite the functions
\[ \frac{1}{4}\Bigl(A(q'-q,p'-p,t)\rho(q'-q,p'-p,t)+ \]
\[ +A(q'+q,p'-p,t)\rho(q'+q,p'-p,t)+ \]
\[ +A(q'-q,p'+p,t)\rho(q'-q,p'+p,t)+ \]
\[ +A(q'+q,p'+p,t)\rho(q'+q,p'+p,t) \Bigr) \]
in the simple form
\[ T_q T_p (A(q,p,t)\rho(q,p,t)) . \]

Let us consider $k$ particle that is described by
generalized coordinates $q_{ks}$ and generalized
momenta $p_{ks}$, where $s=1,...,m$.
The operator $T[k]$ is defined by the relation
$T[k]=T_{q_{k1}} T_{p_{k1}}...T_{q_{km}} T_{p_{km}}$. 
For the $n$-particle system phase space, we use the operator
$T[1,...,n]=T[1]...T[n]$. 

2. Let us define the integral operators $\hat I^{\alpha}_{x_k}$ and
$\hat I^{\alpha}[k]$.
The operator $\hat I^{\alpha}_{x_k}$ is defined by the equation
\be \hat I^{\alpha}_{x_k} f(x_k)=
\int^{+\infty}_{-\infty}  f(x_k) d \mu_{\alpha} (x_k) .\ee
In this case, fractional integral (\ref{FI5}), which defines 
the average value, can be rewritten in the form
\[ <A>_{\alpha}=\hat I^{\alpha}_{x} T_x A(x)\rho(x) . \]

Let us define the phase space integral operator $\hat I^{\alpha}[k]$ 
for $k$ particle by
$\hat I^{\alpha}[k]=
\hat I^{\alpha}_{q_{k1}} \hat I^{\alpha}_{p_{k1}} ...
\hat I^{\alpha}_{q_{km}} \hat I^{\alpha}_{p_{km}}$, 
i.e., we use 
\be \hat I^{\alpha}[k] f({\bf q}_k,{\bf p}_k)=
\int f({\bf q}_k,{\bf p}_k)
d \mu_{\alpha}({\bf q}_k,{\bf p}_k) . \ee
Here $d \mu_{\alpha}({\bf q}_k,{\bf p}_k)$ is an elementary
$2m$-dimensional phase volume that is defined by the equation 
\[ d \mu_{\alpha}({\bf q}_k,{\bf p}_k)=(\alpha \Gamma(\alpha))^{-2m}
d q^{\alpha}_{k1}\wedge dp^{\alpha}_{k1} \wedge ... \wedge
d q^{\alpha}_{km}\wedge dp^{\alpha}_{km} . \]
For the $n$-particle system phase space, we use the integral operator
$\hat I^{\alpha}[1,...,n]=\hat I^{\alpha}[1]...\hat I^{\alpha}[n]$. 

3. Let us define the fractional analog of the average values
$<A>$ for the phase space for $n$-particle system.
Using the suggested notations, we can define
the fractional average value by the relation
\be <A>_{\alpha}=
\hat I^{\alpha}[1,...,n] T[1,...,n] A \rho_{n} .  \ee
In the simple case ($n=m=1$), the fractional average value is defined 
by the equation 
\be \label{AV2} <A>_{\alpha}=
\int^{\infty}_{-\infty} \int^{\infty}_{-\infty}
d\mu_{\alpha}(q,p) \ T_q T_p A(q,p)\rho(q,p) . \ee

Note that the fractional normalization condition \cite{chaos}
is a special case of this definition of the average value
\[ <1>_{\alpha}=1 . \]

\subsection{Reduced Distribution Functions}

In order to derive a fractional analog of the 
Bogoliubov hierarchy equations we must define the
reduced distributions.

Let us consider a classical system with fixed number $n$ of
identical particles. 
Suppose $k$ particle is described by the {\it dimensionless} 
generalized coordinates $q_{ks}$ and generalized
momenta $p_{ks}$, where $s=1,...,m$.
We use the following notations
${\bf q}_k=(q_{k1},...,q_{km})$ and
${\bf p}_k=(p_{k1},...,p_{km})$.

The state of this system can be  described by 
{\it dimensionless} n-particle distribution function $\rho_{n}$
in the $2mn$-dimensional phase space
\[ \rho_{n}({\bf q},{\bf p},t)=
\rho({\bf q}_{1},{\bf p}_{1},...,{\bf q}_{n},{\bf p}_{n},t). \]

We assume that distribution function is invariant under the
permutations of identical particles \cite{Bog2}:
\[ \rho(...,{\bf q}_{k},{\bf p}_{k},...,{\bf q}_{l},{\bf p}_{l},...,t)=
\rho(...,{\bf q}_{l},{\bf p}_{l},...,{\bf q}_{k},{\bf p}_{k},...,t) . \]
In this case, the average values for the classical observables
can be simplified.
We use the tilde distribution functions
\be \tilde \rho_n({\bf q},{\bf p},t)=T[1,...,n]\rho_n({\bf q},{\bf p},t) ,\ee
and the function $\tilde \rho_1$ that is defined by the equation
\be \label{r1} \tilde \rho_{1}({\bf q},{\bf p},t)=
\tilde \rho({\bf q}_{1},{\bf p}_{1},t)=
\hat I^{\alpha}[2,...,n]\tilde \rho_{n}({\bf q},{\bf p},t). \ee
This function is called one-particle reduced distribution function.
The function is defined for the 2m-dimensional phase space.
Obviously, that one-particle distribution function satisfies
the normalization condition \cite{chaos}
\be \label{r3} \hat I^{\alpha}[1] \tilde \rho_{1}({\bf q},{\bf p},t)=1
. \ee

Two-particle reduced distribution function $\tilde \rho_2$
is defined by the fractional integration of the $n$-particle
distribution function over all ${\bf q}_{k}$ and ${\bf p}_{k}$,
except $k=1,2$:
\be \label{p2} \tilde \rho_{2}({\bf q},{\bf p},t)=
\tilde \rho({\bf q}_{1},{\bf p}_{1},{\bf q}_{2},{\bf p}_{2},t)=
\hat I^{\alpha}[3,...,n] \tilde \rho_{n}({\bf q},{\bf p},t) . \ee

\section{Fractional Liouville and Bogoliubov equations}

\subsection{Fractional Liouville Equation}

The fractional generalization of the Liouville equation 
is derived in Ref. \cite{chaos}.
Let us consider the Hamilton equations for n-particle system in the form
\be \label{H3}
\frac{dq^{\alpha}_{ks}}{dt}=G^k_s(q^{\alpha},p^{\alpha}), \quad
\frac{dp^{\alpha}_{ks}}{dt}=AF^k_s(q^{\alpha},p^{\alpha},t). \ee
The evolution of n-particle distribution function $\rho_{n}$
is described by the Liouville equation.
The fractional Liouville equation \cite{chaos} for n-particle distribution 
function has the form
\be \label{L1}
\frac{d \tilde \rho_n}{dt}+ \Omega_{\alpha} \tilde \rho_n =0. \ee
This equation can be derived \cite{chaos} from the fractional
normalization condition 
\be \hat I^{\alpha}[1,...,n] \tilde \rho_{n}({\bf q},{\bf p},t)=1 . \ee
In the Liouville equation $d/dt$ is a total time derivative
\[ \frac{d}{dt}=\frac{\partial}{\partial t}+
\sum^{n,m}_{k,s=1}\frac{dq_{ks}}{dt}\frac{\partial}{\partial q_{ks}}+
\sum^{n,m}_{k,s=1}\frac{dp_{ks}}{dt}\frac{\partial}{\partial p_{ks}} . \]
Using Eq. (\ref{H3}), this derivative can be written for the fractional
powers in the form
\be \label{ttd3} \frac{d}{dt}=\frac{\partial}{\partial t}+
\sum^{n,m}_{k,s=1}
G^k_s\frac{\partial}{\partial q^{\alpha}_{ks}}+
A\sum^{n,m}_{k,s=1}
F^k_s\frac{\partial}{\partial p^{\alpha}_{ks}} . \ee
The $\alpha$-omega function is defined by the equation
\be  \label{o2} \Omega_{\alpha}=\sum^{n,m}_{k,s=1} \Bigl(
\{ G^k_s,p^{\alpha}_{ks}\}_{\alpha}+A
\{q^{\alpha}_{ks},F^k_s\}_{\alpha} \Bigr) . \ee
Here we use the following notations for the fractional Poisson brackets: 
\be \label{PB0} \{A,B\}_{\alpha}=\sum^{n,m}_{k,s=1}\Bigl(
\frac{\partial A}{\partial q^{\alpha}_{ks}}
\frac{\partial B}{\partial p^{\alpha}_{ks}}-
\frac{\partial A}{\partial p^{\alpha}_{ks}}
\frac{\partial B}{\partial q^{\alpha}_{ks}} \Bigr) . \ee

Using (\ref{L1}), (\ref{o2}) and (\ref{ttd3}), we get
the Liouville equation in the form
\be \label{r2}
\frac{\partial \tilde \rho_{n}}{\partial t}=\Lambda_{n} \tilde \rho_{n} , \ee
where $\Lambda_{n}$ is Liouville operator that is
defined by the equation
\be \label{Lam} \Lambda_{n} \tilde \rho_{n} =-
\sum^{n,m}_{k,s=1} \Bigl(
\frac{\partial (G^k_s \tilde \rho_n)}{\partial q^{\alpha}_{ks}}+
A\frac{\partial (F^k_s \tilde \rho_n)}{\partial p^{\alpha}_{ks}}  \Bigr) . \ee

\subsection{First Fractional Bogoliubov Equation}

The Bogoliubov equations  \cite{Bog,Bog3,Petrina,Gur}
are equations for the reduced distribution functions.
These equations can be derived from the Liouville equation.
Let us derive the first fractional Bogoliubov equation (\ref{er1-2})
from the fractional Liouville equation (\ref{r2}).

In order to derive the equation for the function $\tilde \rho_{1}$
we differentiate Eq. (\ref{r1}), which defines 
one-particle reduced distribution:
\[ \frac{\partial \tilde \rho_{1}}{\partial t}=
 \frac{\partial}{\partial t} \hat I^{\alpha}[2,...,n] \tilde \rho_{n}=
\hat I^{\alpha}[2,...,n] \frac{\partial \tilde \rho_{n}}{\partial t} . \]
Using the Liouville equation (\ref{r2}) for
$n$-particle distribution  function $\tilde \rho_{n}$, we have
\be \label{92a} \frac{\partial \tilde \rho_{1}}{\partial t}=
\hat I^{\alpha}[2,...,n] \Lambda_{n} \tilde \rho_{n}({\bf q},{\bf p},t) . \ee
Substituting Eq. (\ref{Lam}) in Eq. (\ref{92a}), we get
\be \label{r1i}
\frac{\partial \tilde \rho_{1}}{\partial t}=- \hat I^{\alpha}[2,...,n]
\sum^{n,m}_{k,s=1} \Bigl(
\frac{\partial (G^{k}_s\tilde \rho_{n})}{\partial q^{\alpha}_{ks}}
+A\frac{\partial (F^{k}_s\tilde \rho_{n})}{\partial p^{\alpha}_{ks}}
\Bigr) . \ee

Let us consider in Eq. (\ref{r1i}) the integration over
$q_{ks}$ and $p_{ks}$ for k-particle term. 
Since the coordinates and momenta are independent variables, 
we can derive
\be \hat I^{\alpha}[q_{ks}]
\frac{\partial }{\partial q^{\alpha}_{ks}} (G^k_s\tilde \rho_{n}) =
\frac{1}{\alpha \Gamma(\alpha)}
\Bigl(G^k_s \tilde \rho_{n} \Bigr)^{+\infty}_{-\infty}=0 . \ee
Here we use the condition
\be \label{lim} \lim_{q_{ks} \rightarrow \pm \infty} \tilde \rho_n =0 , \ee
which follows from the normalization condition.
If limit (\ref{lim}) is  not equal to zero, then
the integration over phase space is equal to infinity.
Similarly, we have
\[ \hat I^{\alpha}[p_{ks}] \Bigl( \frac{\partial}{\partial p^{\alpha}_{ks}}
(F^{k}_s\tilde \rho_{n}) \Bigr) =
\frac{1}{\alpha \Gamma(\alpha)}
\Bigl(F^{k}_s \tilde \rho_{n} \Bigr)^{+\infty}_{-\infty}=0  . \]
Then all terms in Eq. (\ref{r1i}) with $k=2,...,n$
are equal to zero. We have only term for $k=1$.
Therefore Eq. (\ref{r1i}) has the form
\be \label{r1i2} \frac{\partial \tilde \rho_{1}}{\partial t}=
- \sum^m_s\hat I^{\alpha}[2,...,n]\Bigl(
\frac{\partial (G^1_s\tilde \rho_{n})}{\partial q^{\alpha}_{1s}}
+A\frac{\partial (F^{1}_{s} \tilde \rho_{n})}{\partial p^{\alpha}_{1s}}
\Bigr) . \ee

Since the variable ${\bf q}_{1}$ is an independent of
${\bf q}_{2},...{\bf q}_{n}$ and  ${\bf p}_{2},...{\bf p}_{n}$,
the first term in Eq. (\ref{r1i2}) can be written in the form
\[ \sum^m_{s=1} \hat I^{\alpha}[2,...,n]
\frac{\partial (G^1_s \tilde \rho_{n})}{\partial q^{\alpha}_{1s}} =\]
\[ = \sum^m_{s=1}\frac{\partial}{\partial q^{\alpha}_{1s}}
G^1_s \hat I^{\alpha}[2,...,n] \tilde \rho_{n} =
\sum^m_{s=1}
\frac{\partial (G^1_s \tilde \rho_{1})}{\partial q^{\alpha}_{1s}} .\]

The force $F^{1}_s$ acts on the first particle.
This force is a sum of the internal forces
$F^{1k}_s=F_s({\bf q}_{1},{\bf p}_{1},{\bf q}_{k},{\bf p}_{k},t)$, 
and the external force
$F^{1e}_{s}=F^{e}_s({\bf q}_{1},{\bf p}_{1},t)$.
In the case of binary interactions, we have
\be \label{Fie2}
F^{1}_s=F^{1e}_{s}+\sum^{n}_{k=2} F^{1k}_s. \ee

Using Eq. (\ref{Fie2}), the second term in Eq. (\ref{r1i2})
can be rewritten in the form
\[ \hat I^{\alpha}[2,...,n] \Bigl(
\frac{\partial (F^1_s \tilde \rho_{n})}{\partial p^{\alpha}_{1s}}  \Bigr) = \]
\[ =\hat I^{\alpha}[2,...,n] \Bigl(
\frac{\partial (F^{1e}_{s} \tilde \rho_{n})}{\partial p^{\alpha}_{1s}} +
\sum^{n}_{k=2}
\frac{\partial (F^{1k}_{s} \tilde \rho_{n} )}{\partial p^{\alpha}_{1s}}
 \Bigr) = \]
\be  \label{rr1i3} =
\frac{\partial (F^{1e}_{s}\tilde \rho_{1}) }{\partial p^{\alpha}_{1s}} +
\sum^{n}_{k=2}
\frac{\partial}{\partial p^{\alpha}_{1s}} \hat I^{\alpha}[2,...,n]
\Bigl( F^{1k}_{s} \tilde \rho_{n} \Bigr) . \ee
Since the n-particle distribution function $\tilde \rho_{n}$
is a symmetric function for the identical particles, we
have that all $(n-1)$ terms of sum (\ref{rr1i3}) are identical.
Therefore the sum can be replaced by one term with the
multiplier $(n-1)$:
\[  \sum^{n}_{k=2}  \hat I^{\alpha}[2,...,n] \
\frac{\partial}{\partial p^{\alpha}_{1s}} 
\Bigl( F^{1k}_s \tilde \rho_{n} \Bigr) = \]
\[ = (n-1)  \hat I^{\alpha}[2,...,n] \
\frac{\partial}{\partial p^{\alpha}_{1s}}
\Bigl( F^{12}_s \tilde \rho_{n} \Bigr)  . \]
Using $\hat I^{\alpha}[2,...,n]=\hat I^{\alpha}[2]\hat I^{\alpha}[3,...,n]$, 
we have
\[ \hat I^{\alpha}[2] \
\frac{\partial}{\partial p^{\alpha}_{1s}} 
\Bigl( F^{12}_s \hat I^{\alpha}[3,...,n] \tilde \rho_{n} \Bigr) 
=\frac{\partial}{\partial p^{\alpha}_{1s}}
\hat I^{\alpha}[2] \ \Bigl( F^{12}_s \tilde \rho_{2} \Bigr) . \]
Here we use definition (\ref{p2}) of two-particle distribution
function.
Since ${\bf p}_{1}$ is independent of
${\bf q}_{2}$, ${\bf p}_{2}$, we can change the order of
the integrations and the differentiations:
\[ \hat I^{\alpha}[2] \ \frac{\partial}{\partial p^{\alpha}_{1s}}
\Bigl( F^{12}_s \tilde \rho_{2} \Bigr) =
\frac{\partial}{\partial p^{\alpha}_{1s}}
\hat I^{\alpha}[2] F^{12}_s \tilde \rho_{2}  . \]

Finally, we obtain the equation for one-particle reduced
distribution function, 
\be \label{er1-2} \frac{\partial \tilde \rho_{1}}{\partial t}+
\sum^m_{s=1}\frac{\partial (G^1_s \tilde \rho_{1})}{\partial q^{\alpha}_{1s}}
+A\sum^m_{s=1}
\frac{\partial (F^{1e}_s\tilde \rho_{1})}{\partial p^{\alpha}_{1s}} =
(n-1)A I(\tilde \rho_{2}). \ee
Here $I(\tilde \rho_{2})$ is a term with two-particle
reduced distribution function, 
\be \label{I2} I(\tilde \rho_{2})=
- \sum^m_{s=1} \frac{\partial}{\partial p^{\alpha}_{1s}} 
\hat I^{\alpha}[2] F^{12}_s \tilde \rho_{2} . \ee
Equation (\ref{er1-2}) is called a 
{\it first fractional Bogoliubov equation} 
(first equation of Bogoliubov chain).

Let us consider the physical meaning of the term $I(\tilde \rho_{2})$.
The term $I(\tilde \rho_{2})d\mu_{\alpha}({\bf q},{\bf p})$
is a velocity of particle number change in $4m$-dimensional
elementary phase volume
$d\mu_{\alpha}({\bf q}_1,{\bf p}_2,{\bf q}_2,{\bf p}_2)$.
This change is caused by the interactions between particles.
If $\alpha=1$, then we have the first Bogoliubov equation for
non-Hamiltonian systems.

\subsection{Second Fractional Bogoliubov Equation}

The fractional Liouville equation allows us to derive equation for
two-particle reduced distribution function $\tilde \rho_{2}$ in the form
\be \label{er1-4} \frac{\partial \tilde \rho_{2}}{\partial t}=
\sum^{2}_{k=1} \Lambda_{k} \tilde \rho_{2}+
\Lambda_{12} \tilde \rho_{2}+c(n)AI(\tilde \rho_{3}), \ee
where $c(n) =(n-1)(n-2)/2$, and 
$\Lambda_{k}$ is one-particle Liouville operator, 
\[ \Lambda_{k} \tilde \rho_{2} =-
\sum^m_{s=1} \frac{\partial (G^k_s \tilde \rho_{2})}{\partial q^{\alpha}_{ks}}-
A\sum^m_{s=1} \frac{\partial (F^{ke}_s \tilde \rho_{2})}{\partial p^{\alpha}_{ks}} , \]
and $\Lambda_{12}$ is two-particle Liouville operator, 
\[ \Lambda_{12} \tilde \rho_{2}=
A\sum^m_{s=1} \frac{\partial}{\partial p^{\alpha}_{1s}}
\Bigl( F^{12}_s \tilde \rho_{2} \Bigr)+
A\sum^m_s\frac{\partial}{\partial p^{\alpha}_{2s}}
\Bigl( F^{21}_s \tilde \rho_{2} \Bigr), \]
and $I(\tilde \rho_3)$ is a term with the three-particle, 
reduced distribution
\be I(\tilde \rho_{3})=  \sum^m_{s=1}
\hat I^{\alpha}[3] \
\Bigl( \frac{\partial  (F^{13}_s \tilde \rho_{3})}{\partial p^{\alpha}_{1s}}
+ \frac{\partial (F^{23}_s \tilde \rho_{3})}{\partial p^{\alpha}_{2s}}
\Bigr). \ee
The derivation of Eq. (\ref{er1-4})
is the analogous to the derivation of Eq. (\ref{er1-2}).

It is easy to see that Eqs. (\ref{er1-2}) and (\ref{er1-4})
are not closed. The system of equations for the reduced distribution
functions are called the Bogoliubov hierarchy equations.

\section{Fractional Vlasov equation and Debye radius}

\subsection{Fractional Vlasov Equation}

In this section, we derive the fractional analog of the Vlasov 
equation from the first fractional Bogoliubov equation.
Let us consider the particles as statistical independent systems.
In this case, we have
\be \label{2-12}
\tilde \rho({\bf q}_{1},{\bf p}_{1},{\bf q}_{2},{\bf p}_{2},t)=
\tilde \rho({\bf q}_{1},{\bf p}_{1},t)
\tilde \rho({\bf q}_{2},{\bf p}_{2},t) . \ee

Substituting Eq. (\ref{2-12}) in Eq. (\ref{I2}), we get
\[ I(\tilde \rho_{2})=
-\sum^m_s \frac{\partial}{\partial p^{\alpha}_{1s}}
\tilde \rho_{1}[1] \hat I^{\alpha}[2] F^{12}_s \tilde \rho_{1}[2] . \]
Here we use the notation $\rho[k]$ for the distribution function
$\rho({\bf q}_{k},{\bf p}_{k},t)$.
As the result we have the effective forces, 
\[ F^{eff}_s({\bf q}_{1},{\bf p}_{1},t)=
\hat I^{\alpha}[2] F^{12}_s \tilde \rho_{1}[2] . \]
In this case, we can rewrite the term (\ref{I2}) in the form
\be \label{Ir2} I(\tilde \rho_{2})=
- \frac{\partial}{\partial p^{\alpha}_{1s}}
(\tilde \rho_{1} F^{1 eff}_s) . \ee
Substituting Eq. (\ref{Ir2}) in Eq. (\ref{er1-2}), we get
\be \label{p1-1} \frac{\partial \tilde \rho_{1}}{\partial t}+
\sum^m_{s=1}\frac{\partial (G^1_s \tilde \rho_{1})}{\partial q^{\alpha}_{1s}}
+A\sum^m_{s=1}\frac{\partial}{\partial p^{\alpha}_{1s}} \Bigl(
(F^{1e}_{s}+bF^{1 eff}_s) \tilde \rho_{1} \Bigr)=0 , \ee
where $b=n-1$. This equation is a closed equation for one-particle
distribution function with the external force $F^{1e}$
and the effective force $F^{1 eff}$.
Equation (\ref{p1-1}) can be called a fractional Vlasov
equation \cite{Vlasov}.

The fractional Liouville, Bogoliubov and Vlasov equation are better
approximation than its classical analogs for the systems 
with the fractional phase spaces (the fractal dimensional spaces).
For example, the systems that live on some fractals 
(spaces with the fractal dimensions)
can be described by the suggested fractional equations.

\subsection{Debye Radius}

In this section, we consider the Debye radius for the fractional systems
that are defined by the equations 
\be \label{qb} \frac{dq^{\alpha}_{ks}}{dt}=\frac{p^{\beta}_{ks}}{m} , \quad
\frac{dp^{\alpha}_{ks}}{dt}=A F^k_s(q,p) , \ee 
where we use the dimensionless variables $q_k$, $p_k$, $F_k$, $t$. 
Let $r_0=q_0$ be the radius of the interaction,
Here $m=Mr_0/t_0p_0$ is a dimensionless mass, 
where $M$ is a particle mass.
Using $m=1$, we get
\be \label{A2} t_0=\frac{Mq_0}{p_0}, \quad
A=\frac{t_0F_0}{p_0}= \frac{Mq_0}{p^2_0} . \ee
Let us use the condition $p^{2}_0/M=kT_0$ for the characteristic 
momentum $p_0$. 
Note that the condition $p^2/M=kT$ can be realized for 
non-Hamiltonian and dissipative systems \cite{Tar-mplb}. 

The first fractional Bogoliubov equation (\ref{er1-2}) 
for the dimensionless one-particle distribution $\tilde \rho_{1}$
has the following dimensionless form \cite{Sandri}:
\be \label{er1-2dl} 
\frac{\partial \tilde \rho_{1}}{\partial t}+
\sum^3_{s=1}\frac{\partial (p^{\beta}_{1s} 
\tilde \rho_{1})}{\partial q^{\alpha}_{1s}}
+A\sum^3_{s=1}
\frac{\partial (F^1_s\tilde \rho_{1})}{\partial p^{\alpha}_{1s}} =
AB I(\tilde \rho_{2}). 
\ee
The dimensionless first Bogoliubov equation (\ref{er1-2dl}) 
has two characteristic parameters, 
\be 
A=\frac{t_0F_0}{p_0}=\frac{M r_0 F_0}{  p^2_0}=\frac{r_0F_0}{kT_0}
, \quad B=n_0 r^{3\alpha}_0. 
\ee
Let us consider the coefficient $B$. 
It is known \cite{Mand,Schr} that fractal particle system and fractal 
media are described by the power law relation (\ref{Nr}):
\be \label{NrD} n(r) = n_0 r^D , \ee
where $D<3$ and $n_0$ is the $D$-dimensional concentration of 
the $D$-dimensional distribution of particles.
The dimension $D$ of fractal system is connected with order of the 
fractional integrals $\alpha$ by $D=3\alpha$.
The concentration $n_0$ can be defined by the
$D$-dimensional mass density $k$: $n_0=k/ M$ 
which is used in Eqs. (\ref{MR}) and (\ref{Nr}).
To calculate the mass fractal dimension $D$ and concentration $n_0$, 
we can take the logarithm of both sides of  Eq. (\ref{NrD}). 
When we graph $ln(n)$ as a function of $ln(r)$, we have 
\[ ln (n)=D \ ln(r)+ln (n_0) , \]
and we get a value of the fractal dimension  $D$ of fractal media
and parameter $n_0$. 
Therefore these values can be measured for 
homogeneous fractal media.

Let us consider the fractional systems with the force
\be
F_{kl}=\frac{e^2}{4 \pi \varepsilon_0 r^2_0} \frac{1}{|r_k-r_l|^{2\delta}},
\ee
where $r_k$ and $r_l$ are dimensional values of coordinates.
If $\delta=1$, then we have the usual electrostatic interaction.
In this case, the Gauss theorem for the fractional space is not satisfied.
If $2\delta=3\alpha-1$, then the Gauss theorem for the fractional
space is satisfied.
The radius $r_0$ and the force $F_0$ are connected by the equation
\be
F_0=\frac{e^2}{4 \pi \varepsilon_0} \frac{1}{r^{2}_0}.
\ee

Using the relation $AB \sim 1$, we have the
characteristic radius of the interaction
in the fractal media
\be r_0=r_D=\sqrt[3\alpha- 1]{\frac{\varepsilon_0 k T_0}{e^2n_0}} , \ee
which can be called a fractional Debye radius.
If the particle systems or media 
have the integer mass dimension $D=3$, then $\alpha=1$, 
and we get the usual equation for the Debye radius \cite{Debye}.
The fractional radius characterizes the scale $q_0=r_0$ of 
the fractal media or fractal system with
non-integer mass dimension.

\section{Conclusion}

In this paper, we consider the fractional generalizations of the phase volume, 
the phase volume element and the Poisson brackets.
These generalizations lead us to the fractional analog of the phase space. 
The space can be considered as a fractal dimensional space. 
We consider systems on the fractional phase space
and the fractional analogs of the Hamilton equations.
The physical interpretation of the fractional phase space 
is discussed.
The fractional generalization of the average values is derived.

It is known that the fractional derivative of a constant need 
not be zero. This relation leads to the correlation between 
coordinates $q$ and momentum $p$.
Therefore $q$ and $p$ are not independent variables 
in the usual sense.  As the result, the generalization 
of the Poisson brackets with fractional derivatives 
(\ref{PB2}) is not canonical. 
In order to derive equations with fractional derivatives 
we must have a generalization of Darboux theorem  \cite{DNF} for 
symplectic form with fractional exterior derivatives.
However this generalization is an open question at this moment.
In order to define Poisson brackets with the usual relations 
for the coordinates and the momenta we can use 
Poisson brackets (\ref{PB}) 
with the fractional power of coordinates and momenta.

Note that the dissipative and non-Hamiltonian systems can have
stationary states of the Hamiltonian systems \cite{Tarpre}.
Classical dissipative and non-Hamiltonian systems can have
the canonical Gibbs distribution as a solution of the stationary
Liouville equations for this dissipative system \cite{Tar-mplb}.
Using the methods \cite{Tar-mplb}, it is easy to prove
that some fractional dissipative systems can have
fractional analog of the Gibbs distribution (non-Gaussian statistic)
as a solution of the fractional Liouville equations.
Using the methods \cite{Tar-mplb}, it is easy to find
the stationary solutions of the fractional
Bogoliubov equations for the fractional systems.

Note that the quantization of the fractional systems is a 
quantization of non-Hamiltonian dissipative systems. 
Using the method, which is suggested in Refs. \cite{Tarpla1,Tarmsu,Tarsam},
we can realize the Weyl quantization for the fractional systems.
The suggested fractional Hamilton and Liouville equations allow us
to derive the fractional generalization for the quantum systems
by methods suggested in Refs. \cite{Tarpla1,Tarmsu,Tarsam}.

In this paper the fractional analogs of the Bogoliubov hierarchy
equations are derived.
In order to derive this analog we use
the fractional Liouville equation \cite{chaos},
we define the fractional average values and the fractional
reduced distribution functions.
The fractional analog of the Vlasov equation
and the Debye radius are considered.

The fractional Bogoliubov hierarchy equation can be used to derive
the Enskog transport equation. 
The fractional analog of the hydrodynamics equations
can be derived from the first fractional Bogoliubov equation.
These equations will be considered in the next paper.

It is known that the Fokker-Planck equation can be derived
from the Bogoliubov hierarchy equations \cite{Is}.
The fractional Fokker-Planck equation
can be derived from the fractional Bogoliubov equation.
However this fractional Fokker-Plank equation
can be differed from the equation known
in the literature \cite{MK,Zas2,Zas}.

The quantum generalization of the suggested fractional Bogoliubov 
equation can be considered by the methods that 
are suggested in Refs. \cite{Tarpla1,Tarmsu,Tarsam}.


\end{document}